\documentclass[conference]{IEEEtran}
\IEEEoverridecommandlockouts
\usepackage{cite}
\usepackage{amsmath,amssymb,amsfonts}
\usepackage{graphicx}
\usepackage{psfrag}
\usepackage{textcomp}
\usepackage{xcolor}
\usepackage{xspace}
\usepackage{xurl}
\usepackage{siunitx}
\def\BibTeX{{\rm B\kern-.05em{\sc i\kern-.025em b}\kern-.08em
    T\kern-.1667em\lower.7ex\hbox{E}\kern-.125emX}}
\usepackage{mathtools}
\usepackage{multirow}
\usepackage{adjustbox}
\usepackage{caption}
\usepackage{subcaption}
\usepackage[french,boxed,vlined,linesnumbered,inoutnumbered,rightnl,algo2e]{algorithm2e}
\usepackage{algorithm}
\usepackage{mwe}
\usepackage{stfloats}
\usepackage{pifont}
\usepackage[table]{xcolor}
\usepackage{tikz}

\newcommand{\blinded}[1]{#1}

\newcommand{\SmaTable}{\textit{Smatable} }

\begin{document}
\title{Towards an End-To-End System for Real-Time Gesture Recognition from Surface Vibrations}
\author{
     \IEEEauthorblockN{
     \blinded{Florian~Hettstedt$^{1,2}$}
     \blinded{Cedric~Giese$^{1}$}
     \blinded{Tianheng~Ling$^{1,2}$}
     \blinded{Keiichi~Yasumoto$^{3,4}$}
     \blinded{Gregor~Schiele$^{1,2}$}
     \blinded{Andreas~Erbslöh$^{1,2}$}}
     \IEEEauthorblockN{\blinded{$^1$Intelligent Embedded Systems Laboratory, University of Duisburg-Essen, Germany}}
     \IEEEauthorblockN{\blinded{$^2$PALUNO, The Ruhr Institute for Software Technology, Essen, Germany}}
     \IEEEauthorblockN{\blinded{$^3$Ubiquitous Computing Systems Laboratory, Nara Institute of Science and Technology, Nara, Japan}}
     \IEEEauthorblockN{\blinded{$^4$RIKEN, Center for Advanced Intelligence Project, Tokyo, Japan}}     
}
\maketitle

\begin{abstract}
Sensing surface vibrations promise unobtrusive interaction for smart home systems by enabling gesture recognition on existing everyday surfaces without disturbing living-space design. 
Existing approaches typically address only parts of the processing chain, such as sensing hardware or offline gesture recognition, rather than providing an end-to-end system from surface-mounted sensors to the evaluation of the prediction model. 
This paper presents a custom sensor system and a configurable data-to-model pipeline for gesture recognition on a standard office desk. Our hardware enables a low-noise sensing of the vibrations using piezoelectric sensors. Building on a modular signal-processing framework, we model the full chain from continuous recordings through variable pre-processing to a model-ready dataset, and process the resulting data with compact depthwise separable 1D-CNNs. We conduct a joint search over pre-processing and model hyperparameters and identify a configuration with~8,722~parameters that uses band-pass filtering, fixed-length windows, and min-max normalization. On a self-recorded dataset with~15~participants performing six~gestures this configuration achieves high accuracies across different data splitting methods, including strong user-independent performance in a leave-one-subject-out cross-validation. 
\end{abstract}


\section{Introduction} 
\label{sec:introduction}

In recent years, smart-home systems have become widely used in many households, but everyday interaction is still dominated by voice assistants and touch displays, which can raise privacy concerns, show inconsistent reliability in real-world conditions, and often integrate poorly with interior aesthetics. This motivates natural, unobtrusive alternatives such as smart furniture, where material surfaces become user interfaces while retaining their primary function and appearance. To avoid the drawbacks of current smart-home interfaces, smart furniture systems should emphasize local, event-driven processing, seamless integration into living spaces, and non-destructive, retrofit-friendly installation, supported by low-power sensing hardware and on-device machine learning.

Yoshida et al.~\cite{Yoshida20232} proposed \SmaTable, which utilizes multiple hidden vibration sensors, custom analog front-ends, and deep learning techniques for on-surface gesture recognition on a table with swipes as the reference task. While the results are promising and demonstrate the feasibility of hidden vibration-sensing for gesture recognition, the evaluation relies on a small self-recorded dataset ($n$=3 participants), a comparatively extensive hardware setup, and offline processing on pre-segmented windows with dataset-specific pre-processing, rather than continuous, on-device operation. This leads to a mismatch between training conditions and the requirements in a real-time embedded system.

This paper presents an approach to design an end-to-end signal processing pipeline for vibration-based gesture recognition in real-time as a basis for translation into embedded hardware. This includes (i) the definition of an end-to-end system combining low-noise sensing hardware, pre-processing, and inference for continuous data streams; and (ii) the optimization search for finding the best configuration. Our main contributions are as follows:
\begin{itemize}
    \item We present a custom all-in-one hardware for sensing surface vibrations with suitable embedded integration for pre-processing and deep learning inference for gesture recognition~(Section~\ref{sec:system_overview}-A).
    \item We integrate data acquisition on-device and evaluate a deep learning-based signal processing pipeline (data-to-model pipeline) that enables event-based classification from continuous data streams~(Section~\ref{sec:system_overview}-B).
    \item We demonstrate the performance of the system on a standard furniture surface by creating and evaluating a self-recorded dataset. Here, a study is done with 15~people by capturing six different gestures~(Section~\ref{sec:dataset-recording}).
    \item We evaluate the end-to-end performance by optimizing the configuration of data-to-model pipeline and adapting the deep learning architecture to achieve the best accuracy or highest resource-efficiency~(Section~\ref{sec:evaluation}).
\end{itemize}
\begin{figure*}[!t] \centering
    \vspace{-10pt}  
    \psfrag{U0}[c][c]{\scriptsize{$V_\mathrm{sns}$ }}
    \psfrag{R0}[c][c]{\scriptsize{$R_\mathrm{sns}$ }}
    \psfrag{U1}[l][l]{\scriptsize{$V_\mathrm{in}$ }}
    \psfrag{R1}[c][c]{\scriptsize{$R_\mathrm{L}$ }}
    \psfrag{R2}[c][c]{\scriptsize{$R_\mathrm{F1}$ }}
    \psfrag{R3}[c][c]{\scriptsize{$R_\mathrm{F0}$ }}
    \psfrag{C0}[l][l]{\scriptsize{$C_\mathrm{F0}$ }}
    \psfrag{C1}[c][c]{\scriptsize{$C_\mathrm{F1}$ }}
    \psfrag{C2}[c][c]{\scriptsize{$C_\mathrm{D0}$ }}
    \psfrag{R4}[c][c]{\scriptsize{$R_\mathrm{D0}$ }}
    \psfrag{C3}[l][l]{\scriptsize{$C_\mathrm{D1}$}}
    \psfrag{R5}[l][l]{\scriptsize{$R_\mathrm{D1}$}}
    \psfrag{R6}[l][l]{\scriptsize{$R_\mathrm{A0}$ }}
    \psfrag{C4}[c][c]{\scriptsize{$C_\mathrm{A0}$ }}
    \includegraphics[width=0.84\linewidth]{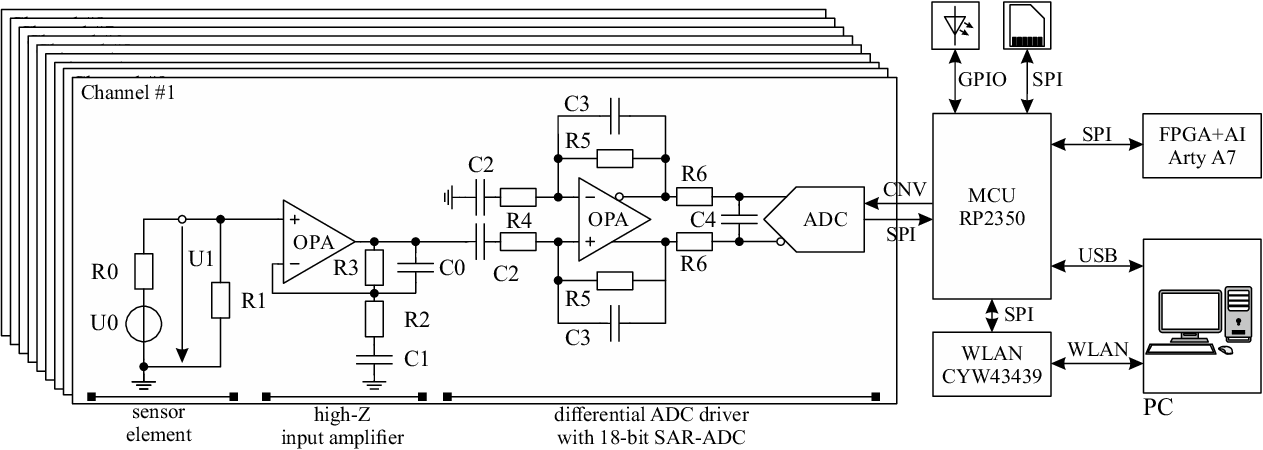}
    \includegraphics[width=0.325\linewidth]{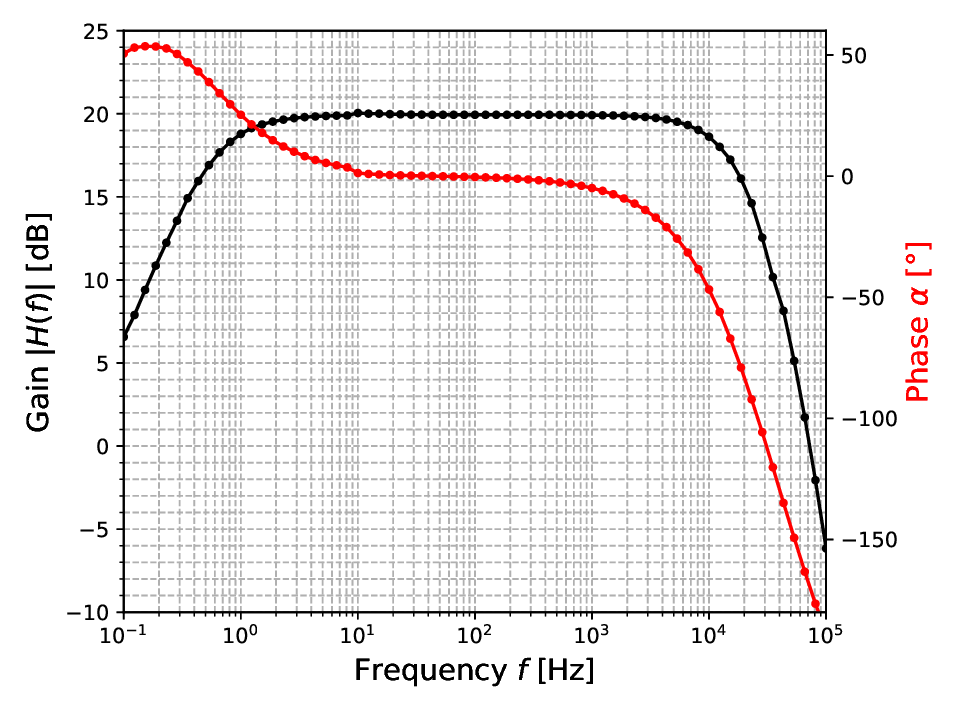}
    \includegraphics[width=0.325\linewidth]{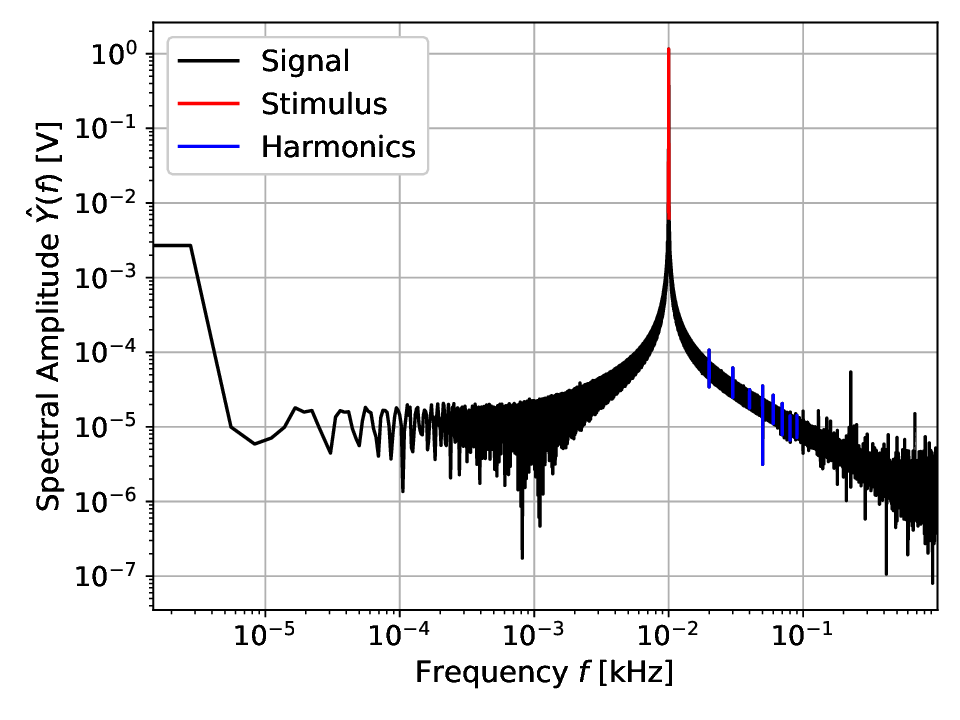}
    \includegraphics[width=0.325\linewidth]{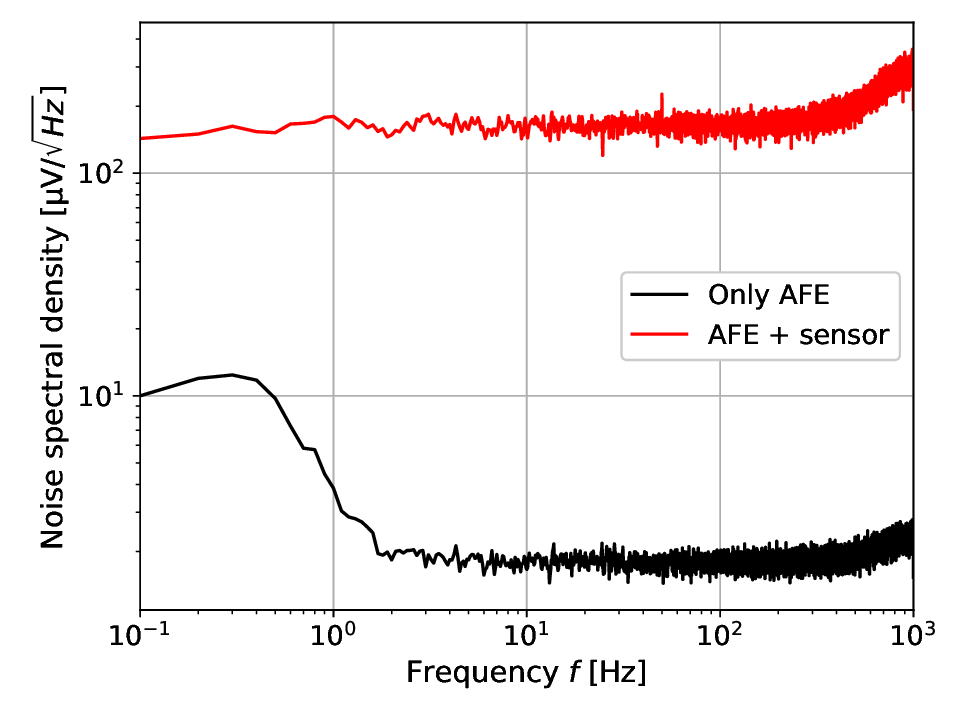}
    \caption{Top: Hardware concept including circuit schematic from one channel of the 9-channel AFE setup, Down: (left) results from the frequency response analysis from AFE input to analog-digital-converter~(ADC) input, (middle) results from the total harmonic distortion analysis at the ADC input, (right) results from the noise transient measurement.}
    \label{fig:setup_afe}
    \vspace{-16pt}  
\end{figure*}
\section{Related Work}  \label{sec:related_work}
Previous work on-surface interaction has explored optical (\textit{HIC}~\cite{8401374}), capacitive (\textit{Sensurfaces}~\cite{10.1145/3534616}), acoustic, and vibration-based sensing approaches.
\textit{HapTable}~\cite{HapTable}, \textit{Sprayable User Interfaces}~\cite{Wessely2020}, and \textit{Smart Table Surface}~\cite{SmarttableDining} track user actions or recognize touch and hand poses on interactive tabletops and smart furniture, but rely on visible sensor units that are difficult to retrofit into existing surfaces.

To reduce instrumentation, acoustic- and vibration-based approaches exploit sound and vibration waves generated by interactions with the surface. \textit{Acustico}~\cite{10.1145/3379337.3415901} uses wrist-worn passive acoustic sensors to detect tap events and estimate their positions on uninstrumented surfaces, and \textit{SAWSense}~\cite{10.1145/3544548.3580991} repurposes voice pickup units and a machine learning pipeline to classify taps, scratches, and swipes on everyday objects. Related work further demonstrates acoustic and vibration-based input for touch and handwriting recognition, like \textit{VibSense}~\cite{VibSense}, \textit{Wrist Powered Touch}~\cite{WristPoweredTouch}, and \textit{WritingRecorder}~\cite{SprayableUserInterface}.
These systems show that acoustic and vibration signals can support rich input vocabularies, but they require wearable sensors, direct contact with the interaction region, or specific hardware, which limits an unobtrusive integration into furniture.

Without affecting the furniture's design, \SmaTable~{\cite{Yoshida20231, Yoshida20232}} mounts piezoelectric sensors in hidden locations on the underside of a tabletop and uses a spectrogram-based 2D convolutional neural network (CNN) to classify swipe gestures.
Building on this line of work, Shibata et al.~\cite{Shibata2025} revisit the \SmaTable datasets and target embedded deployment by replacing spectrogram features and 2D-CNNs with 1D-CNNs and depthwise separable 1D-CNNs on downsampled waveforms combined with integer-only quantization, enabling FPGA-based inference. Together, these works show that vibration-based gesture recognition on furniture and embedded inference are feasible. 
However, they are still based on a small dataset ($n=3$), rely on a bulky sensing hardware with offline and dataset-specific pre-processing pipelines, and do not address the co-design of sensing hardware and continuous data-to-model processing.
Motivated by these advances, we present a system that combines hidden vibration sensing, a custom all-in-one hardware platform, and a configurable data-to-model pipeline to enable an end-to-end embedded deployment.
\section{System Design} \label{sec:system_overview} 
In this section, we describe the (i) requirements, concept and characteristics of the all-in-one hardware design for sensing gesture behavior and the (ii) digital end-to-end digital signal processing pipeline in order to recognize the gesture.
\subsection{Hardware Design}
To design the analog front-end~(AFE), we have to fulfill the application-specific requirements. In our scenario, piezoelectric sensors (Murata 7BB-41-2L0,1) are placed onto the table surface using strong magnets to sense the vibrations, like shown in Figure~\ref{fig:desktop} (bottom). Each element is connected using passive-shielded SMA cable to one AFE channel and it returns an electrical voltage~$V_\mathrm{sns}$ that depends on the mechanical expansion of the piezo-crystal. These sensors have a high internal resistance that requires a high input impedance of the AFE ($>$\SI{10}{M\Omega}). For detecting swipes on a table, the sensor signal strength is weak, which requires a low-noise design (input noise $<$ \SI{100}{\micro V_{rms}}). 

\begin{figure}[t]
    \centering
    \includegraphics[width=0.98\columnwidth]{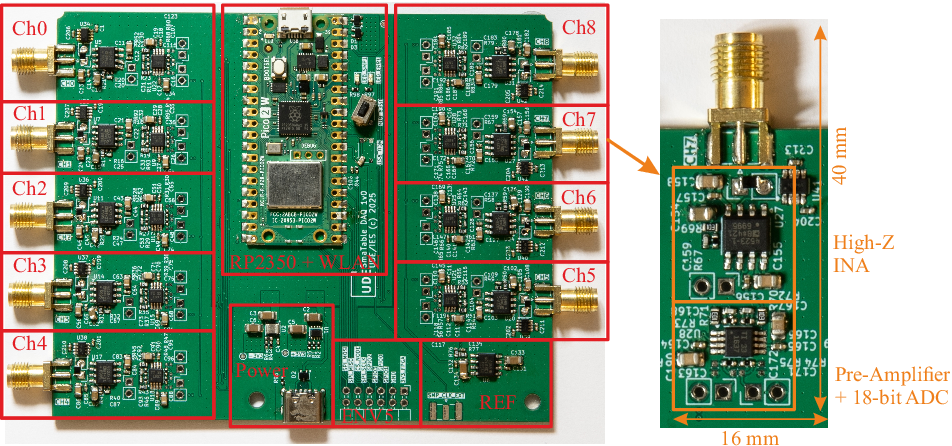}
    \caption{Image of the assembled hardware design [\SI{110}{mm}$\times$\SI{90}{mm}] with a zoom on channel of the analog front-end from~Figure~\ref{fig:setup_afe}.}
    \label{fig:hardware_board}
\end{figure}

These constraints result in the hardware concept, shown in Figure~\ref{fig:setup_afe} (top) with the assembled hardware in Figure~\ref{fig:hardware_board}. The hardware enables simultaneous sensor data acquisition up to nine channels. 
Each of the channels consists of three stages. 
The sensor element is connected to a resistive load in order to avoid a current flow into the first amplifier stage. 
Afterwards, the sensor signal~$V_\mathrm{in}$ is captured by a high-impedance input amplifier and pre-amplified using a differential analog-digital converter~(ADC) drive-to-drive the full dynamic range of the 18-bit SAR-ADC. Both stages includes a band-pass filter characteristic. 
The data acquisition is triggered continuously from the microcontroller (Raspberry~Pi~RP2350A). The raw data stream passes the embedded signal processing pipeline and the output is transmitted to a computer using the USB or WLAN interface. Due to the interface data bandwidth, a maximum sampling rate of~\SI{12.5}{kHz} for transmitting the raw data of all channels is available. The system has a power dissipation of~\SI{226.3}{mW}@\SI{5}{V} and can be powered via a computer or a power bank using the USB connector. 
The deep neural network to predict gestures can be deployed on the MCU or on an external FPGA board.
\begin{table}[h]
    \caption{Overview of the hardware metrics of the proposed hardware ($n\!=\!9)$ compared with the \SmaTable system in ~\cite{Yoshida20231}. }
    \label{tab:hardware_metric}
    \resizebox{\columnwidth}{!}{%
    \begin{tabular}{|l|c|c|c|} \hline
        \textbf{Metric} & \textbf{Unit} & \textbf{Yoshida et al. ~\cite{Yoshida20231}} & \textbf{Our hardware} \\\hline\hline
        Midband gain                & dB    & \SI{18}{}         & \SI{20.01}{}$\pm$\SI{0.16}{}\\ \hline
        Midband gain                & V/V   & \SI{8}{}          & \SI{10}{}\\ \hline
        High-pass corner freq.      & Hz    & n.a.              & \SI{0.5}{}$\pm$\SI{0.08}{}\\ \hline
        Low-pass corner freq.       & kHz   & \SI{30}{}         & \SI{21.4}{}$\pm$\SI{0.89}{}\\ \hline
        Total harmonic distortion   & dB    & n.a.              & \SI{103.2}{}$\pm$\SI{2.98}{}\\ \hline
        Eff. input noise (AFE) & \SI{}{\micro V_{rms}} & \SI{26.64}{}*  & \SI{6.65}{}$\pm$\SI{0.25}{}\\ \hline
        Eff. input noise (sensor) & \SI{}{\micro V_{rms}} & n.a.       & \SI{110.5}{}$\pm$\SI{0.28}{}\\\hline
    \end{tabular}
    }
    *extracted the signal-to-noise ratio
\end{table}
Table~\ref{tab:hardware_metric} summarizes the extracted hardware metrics like (i) gain and filter characteristic, (ii) harmonic distortion and (iii) noise properties. The measurement protocols and results are explained below.

Figure~\ref{fig:setup_afe}~(down, left) shows the gain profile of an AFE channel which is measured with the frequency response analysis~(FRA). Here, a sinusoidal waveform (\SI{20}{mV_{pp}}, \SI{0.1}{Hz} to~\SI{100}{kHz}) is applied to the AFE input and the output of the ADC driver is feedbacked into the FRA device. In addition, Figure~\ref{fig:setup_afe} (down, middle) shows the FFT spectrum output for determining the harmonic distortion. This is done by applying a sinusoidal signal (\SI{200}{mV_{pp}}, \SI{10}{Hz}) to the AFE input, which is generated from the R\&S~MXO44, and capturing the ADC raw data stream for~\SI{300}{s} at a sampling rate of~\SI{2}{kHz}. For calculating the total harmonic distortion~(THD), the ratio of the square sum of all harmonic amplitudes (highlighted in blue) and the stimulus amplitude (highlighted in red) is calculated. 


In Figure~\ref{fig:setup_afe} (bottom, right), the output noise spectral density of the AFE is shown. The spectrum is calculated from the transient measurement by (i) shorting the input to system ground and (ii) connecting a sensor element to the input. In both cases, the ADC data stream is captured for a time window of~\SI{300}{s} at a sampling rate of~\SI{2}{kHz}. It is important to note, that the noise properties of the sensor is dominating with the thermal noise behavior and an output noise density of~\SI{180.3}{\micro V/\sqrt{Hz}}. The noise level of the AFE is decreased by a factor of~\SI{100} with an output noise density of~\SI{18.34}{\micro V/\sqrt{Hz}}. The effective input noise in Table~\ref{tab:hardware_metric} defines the minimal sensor level to achieve a signal-to-noise ratio of~1 by integrating the spectral noise density and divide it by the gain profile. 

In addition, we compared our solution with the hardware metrics from the original paper and we achieve better noise properties for similar design parameters (gain, filter).
\subsection{End-to-End Signal Processing Pipeline}
Prior work on end-to-end pipeline design for embedded deep learning highlights that signal pre-processing, model architecture, and hardware constraints must be co-designed. The~\textit{denspp.offline} framework~\cite{Buron2023} provides a Python framework for evaluating signal-processing pipelines in continuous data-stream applications, by emulating the chain from an analog front-end through digital signal pre-processing, feature extraction, and classification as composable module blocks. Different pipeline variants can be benchmarked for prediction performance and latency to expose the total computational footprint before committing to a hardware implementation.
\begin{figure}[ht] \centering
    \includegraphics[width=0.9\linewidth]{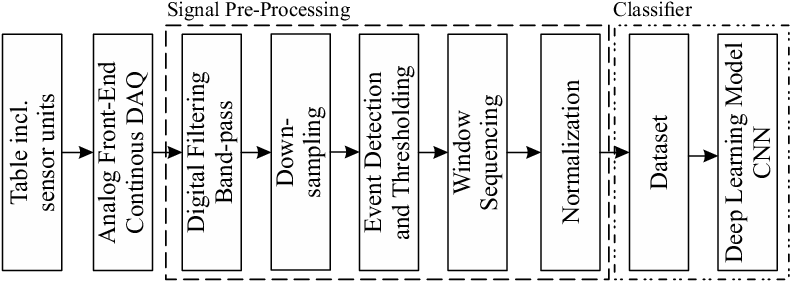}
    \caption{Workflow of the configurable data-to-model pipeline.}
    \label{fig:data-to-model-pipeline}
\end{figure}
\subsubsection{Data-to-Model Pipeline}
We transform the continuous ADC data stream into event-based windows as input to deep learning models using a configurable data-to-model pipeline implemented with the \textit{denspp.offline} framework~\cite{Buron2023}. The pipeline is composed of modular blocks whose order and parameters can be adjusted, which enables a systematic exploration of different preprocessing variants in a hyperparameter search.
Figure~\ref{fig:data-to-model-pipeline} summarizes the signal processing pipeline from the sensor units to the dataset used for model training. A band-pass filter block can be applied either to the continuous data stream or to individual windows. We use a second-order IIR band-pass filter of Butterworth type, with configurable cut-off frequencies. An optional event detection block marks candidate gesture intervals  based on a configurable thresholding scheme.
Around these intervals, a windowing block extracts fixed-length windows with a configurable length, so that different temporal contexts can be investigated. If event detection is disabled, gesture events are annotated manually and the windowing operates directly on these labeled intervals.
A normalization block maps each window to a fixed value range using a configurable normalization method. An additional downsampling block can reduce the effective sampling rate by an integer factor, decreasing the dimensionality of each window. 
By treating the filter band, the window length, normalization method, and downsampling factor as pipeline parameters, we obtain different preprocessed versions of the dataset that can be combined with different model configurations in the hyperparameter search.
\begin{figure}[ht]    \centering
    \includegraphics[width=0.92\columnwidth]{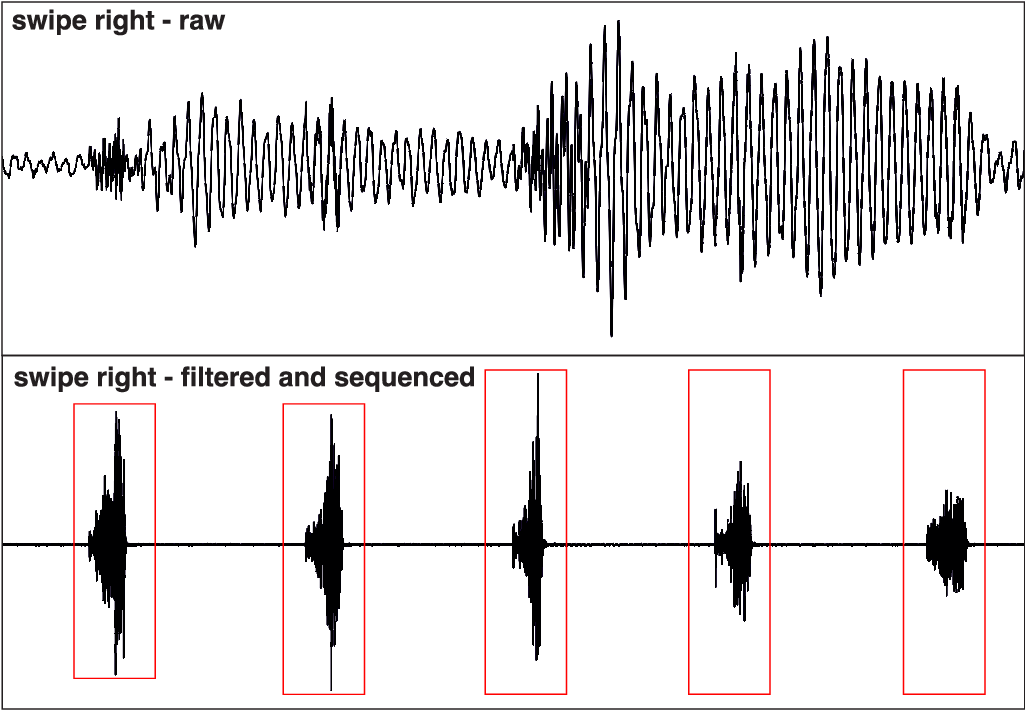}
    \caption{Signal of a swipe right gesture on one channel in raw form and after pre-processing (including band-pass filtering, event detection, and window sequencing).}
    \label{fig:raw-vs-filtered-signal}
\end{figure}
Figure~\ref{fig:raw-vs-filtered-signal} shows an example gesture on one channel before and after applying band-pass filtering, event detection, and window sequencing.
\subsubsection{Deep Learning Model}
Recent work~\cite{Shibata2025} has shown that 1D-CNNs and 1D-SepCNNs achieve high accuracies on the original \SmaTable \cite{Yoshida20232} dataset and can be deployed into an embedded system.
Following this, we adopt a similar 1D-SepCNN architecture to our sensing setup and data-to-model pipeline. In preliminary experiments, plain 1D-CNN variants exhibited stronger overfitting and lower validation accuracy than 1D-SepCNNs, so we focus on the latter in the following.
\begin{figure}[ht]    
\centering
    \includegraphics[width=0.65\linewidth]{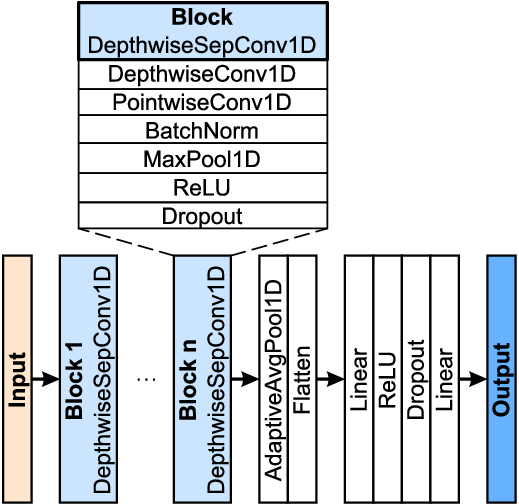}
    \caption{1D-SepCNN architecture for vibration-based gesture recognition with an adaptable number of convolutional blocks.}
    \label{fig:1d-sep-cnn-architecture}
\end{figure}
Figure~\ref{fig:1d-sep-cnn-architecture} illustrates the resulting model. Given a preprocessed gesture window, the model applies a stack of depthwise separable convolutional blocks, each consisting of a 1D-SepCNN with one filter per input channel followed by a pointwise 1D-convolution with a kernel size of~1, batch normalization, max-pooling with a kernel size of~2, ReLU activation and dropout. All depthwise layers share the same kernel size. The kernel size, the number of output channels per block, and the number of blocks are treated as hyperparameters. After the last block, adaptive average pooling and a flattening operation reduce the time dimension, followed by a small two-layer fully connected classifier that maps to the gesture classes.
 
\section{Dataset Recording}
\label{sec:dataset-recording}
To evaluate our end-to-end design, we recorded a new dataset using our all-in-one hardware platform. The dataset contains vibration signals for the four swipe gestures (left, right, up, down) from the original \SmaTable dataset~\cite{Yoshida20231, Yoshida20232} and two additional gestures (tap and knock). 
This section describes (i) the recording setup and sensor placement, (ii) the data acquisition procedure and (iii) the semi-automatic labeling procedure used to label gesture onsets.
\subsection{Sensor Setup}
\begin{figure}[ht]  \centering
    \psfrag{C0}[c][c]{\small{CH1}}
    \psfrag{C1}[c][c]{\small{CH4}}
    \psfrag{C2}[c][l]{\small{CH8}}
    \psfrag{C3}[c][c]{\small{CH6}}
    \psfrag{M0}[c][c]{\small{\SI{39}{mm}}}
    \psfrag{M1}[c][c]{\small{\SI{66}{mm}}}
    \includegraphics[width=0.68\linewidth]{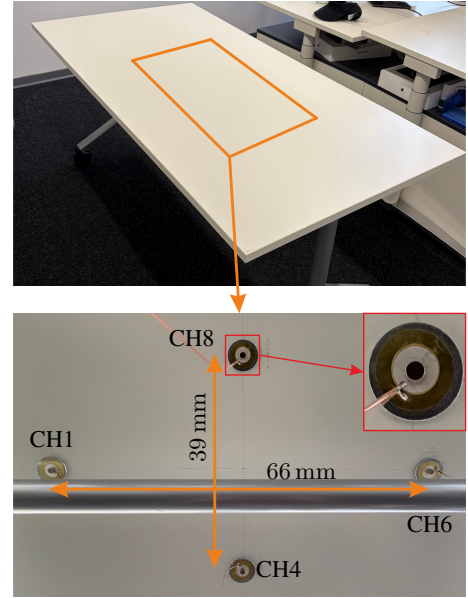}
    \caption{(Top) Top view of the interaction table (\SI{160}{cm} $\times$ \SI{80}{cm}). 
    (Bottom) Bottom view with sensor placement and a zoomed-in sensor unit.}
    \label{fig:desktop}
\end{figure}

The sensor and recording setup follows the~\SmaTable\cite{Yoshida20231, Yoshida20232} configuration which is shown in Figure~\ref{fig:desktop}. Four sensor units were mounted on the underside of an office desk. Each unit consists of a self-adhesive metal disc, a piezoelectric sensor, and a neodymium magnet, which fixes the unit to the desk and acts as a seismic mass to increase sensitivity. Two units are placed along an Y-axis and two along an X-axis, forming an orthogonal cross. The sensors are connected via shielded cables to the custom hardware, which streams the signals with a sampling rate of~\SI{1}{kHz} per channel to a laptop, where they are stored in binary~\textit{XDF} format.
\subsection{Data Acquisition}
The dataset was collected from 15~participants performing six gestures: swipe left, swipe right, swipe up, swipe down, tap, knock. Gestures were carried out on the tabletop of a height-adjustable desk measuring~\SI{160}{cm}~$\times$~\SI{80}{cm} with a height of~\SI{71.5}{cm}. The desk had a~\SI{25}{mm} three-layer fine chipboard top with a melamine resin coating and a transverse crossbeam between the legs.
All gestured except knock were performed with one finger; knock was executed with the knuckles. Participants used the same finger and hand throughout. Gestures were performed at the geometric center of the tabletop, aligned with the center of the four sensor layout. Participants stood in front of the desk in a fixed position. 
For each participant, we recorded ten~sessions. Each session comprised~ten~repetitions of each of the~six~gestures, resulting in~60~events per session, ~600~events per participant, and 9,000~events in total. Each session was recorded and stored independently. Gesture strength and length were left to the participants' discretion to capture intra-participant variability. Participants were instructed to minimize extraneous movement and speech and to avoid incidental contact with the desk between gestures.
\subsection{Semi-automatic Event Annotation}
To reduce the manual effort for annotating gesture onsets in our self-recorded dataset, we adapt our processing pipeline to extract the event timestamps in a semi-automatic manner.

We configure a reduced event-detection pipeline that operates on each recording file. First, a band-pass filter (fixed lower and upper cut-off frequencies of~\SI{225}{Hz} and~\SI{375}{Hz}) is applied to the multi-channel data. From the filtered stream, we derive an one-dimensional signal by computing the mean of absolute values across channels at each time point. On this 1D signal, we compute the median absolute derivation and define two threshold values: a high threshold with a gain of~4 and a low threshold with a gain of~2. Event candidates are then detected using a sliding window. For each window, we compute its peak value and the proportion of samples that lie above the low threshold. A window is marked as containing a gesture if its peak exceeds the high threshold and at least~\SI{40}{\%} of its samples are above the low threshold. To avoid repeated triggering by the same gesture, the detector is locked for a fixed time interval after each event.

Instead of producing model-ready windows, this pipeline outputs a time-sorted list of event timestamps for each recording session and task. We manually refine this list by removing false positives and adding potential missing events. Because the candidate timestamps are already closely aligned with the actual gesture occurrences, the manual work is reduced to checking and correcting a small number of cases. The final timestamps are then used as event onsets for window sequencing in the data-to-model pipeline. In practice, out of 900 recording files (one per task, session, participant), 824~(\(\approx 91.6\%\)) could be annotated fully automatically, and only 76 files required manual corrections. Based on this annotated dataset, we next optimize and evaluate the end-to-end data-to-model pipeline.

\section{Evaluation and Discussion} \label{sec:evaluation}
In this section, (i) we optimize the configuration for the data-to-model pipeline, (ii) evaluate its performance using different data splitting methods, and (iii) compare the results to prior work.
\subsection{Pre-processing and Model Configuration Search}
The pre-processing pipeline and the 1D-SepCNN architecture define our data-to-model configuration space. We first restrict this space using design considerations and preliminary experiments, discarding consistently underperforming choices such as extremely small or overly large models, omitting normalization, or very low and narrow band-pass ranges. On the reduced space, we then run a systematic grid search over combinations of pre-processing blocks and model hyperparameters as introduced in Section~\ref{sec:system_overview}-B. Table~\ref{tab:hyperparam} lists the resulting hyperparameter ranges.
\begin{table}[ht]   
\centering
\caption{Search space of pre-processing, model, and training hyperparameters used in the joint configuration search.}
\label{tab:hyperparam}
\resizebox{\columnwidth}{!}{%
\begin{tabular}{|l|l|l|l|}
\hline
\textbf{Parameters}         & \textbf{Values}                               & \begin{tabular}[c]{@{}c@{}}\textbf{Best} \\ \textbf{Accuracy}\end{tabular}  & \begin{tabular}[c]{@{}c@{}}\textbf{Least} \\ \textbf{Parameter}\end{tabular} \\ \hline\hline

Band-pass filter range      & \{none, 225-375, 300-450\}  & 225 - 375\  & 225 - 375\\ \hline 
Downsampling ratio          & \{1 (none), 2, 5, 10\}                        & 1 & 2\\ \hline
Window length [ms]          & \{1,000, 1,250, 1,500\}                & 1,250 & 1,500 \\ \hline
Kernel size                 & \{9, 15, 25, 33, 39\}             & 15 & 9 \\ \hline
Blocks $\times$ channel width      & \{4$\times$16, 4$\times$32, 6$\times$16, 6$\times$32\} & 6$\times$32 & 4$\times$16 \\\hline
Dropout                     & \{0.2, 0.3\}                             & 0.2 & 0.2 \\\hline
\end{tabular}
}
\end{table}

For each candidate configuration, we instantiate the pre-processing pipeline and 1D-SepCNN model and train it on our self-recorded dataset using 5-fold cross-validation. Folds are defined by an 80/20 split of recording sessions for each participant (eight sessions for training, two for testing), and sessions from all participants are pooled to train a single model. We train each configuration for~300~epochs with a batch size of~128, min-max normalization, and the AdamW optimizer with a learning rate of~1e-3. Configurations are selected based on mean test accuracy across folds and, in case of similar performance, on a smaller number of parameters.

After running the grid search, the selected configuration corresponds to the pre-processing and model hyperparameters in Table~\ref{tab:hyperparam} and yields a model with~8,722 parameters and an accuracy of~\SI{97.32}{\%} across all six gestures. For comparison, the smallest evaluated model achieves~\SI{91.61}{\%} accuracy with only~1,802 parameters.
\subsection{Performance on Different Data Splitting Methods}
In the following, we want to evaluate the system using three data splitting schemes which are similar to the original~\SmaTable approach~\cite{Yoshida20231, Yoshida20232}.
In the \textit{Per Subject~(PS)} setting, we train and evaluate each participant independently using cross-validation on the session-level. Since our dataset contains~10~sessions per participant (instead of nine), we use~5-fold cross-validation so that all sessions are distributed evenly across folds. Applying \textit{Leave-One-Subject-Out~(LOSO)}, all sessions of one participant form the test set and the sessions of the remaining participants form the training set, yielding~15~folds. The \textit{Add-One-Session~(AOS)} settings extend \textit{LOSO} by adding one calibration session from the held-out participant to the training set and using the remaining sessions of that participant for testing. We use the first session of each participant as the calibration session.
\begin{table}[ht]   
\centering
\caption{System-level comparison of classification accuracy between the original~\SmaTable system~\cite{Yoshida20232}, the improved model~\cite{Shibata2025}, and our proposed system across three data splitting methods with the best configuration.}
\label{tab:comparison-ford}
\resizebox{1\columnwidth}{!}{
\begin{tabular}{|c|c|c|c|c|}
\hline
\multirow{2}{*}{\begin{tabular}[c]{@{}c@{}}\textbf{Split} \\ \textbf{Method}\end{tabular}} & \multirow{2}{*}{\textbf{System}} & \multirow{2}{*}{\begin{tabular}[c]{@{}c@{}}\textbf{Num. of} \\ \textbf{Gestures}\end{tabular}} & \multicolumn{2}{c|}{\textbf{Evaluation Metrics}} \\ \cline{4-5}
& & & \textbf{Accuracy} & \textbf{Precision} \\\hline\hline

\multirow{4}{*}{PS} & 
Yoshida et al.~\cite{Yoshida20232}    & 4   & 0.920  & n.a.  \\ \cline{2-5}
& Shibata et al.~\cite{Shibata2025}   & 4   & 0.970  & n.a.  \\ \cline{2-5}
& \multirow{2}{*}{Our Approach}       & 4   & $0.941 \pm 0.063$ & $0.949 \pm 0.052$  \\  \cline{3-5}
&                                     & 6   & $0.961 \pm 0.045$ & $0.965 \pm 0.038$  \\\hline\hline

\multirow{4}{*}{LOSO} & 
Yoshida et al.~\cite{Yoshida20232}    & 4   & 0.670  & n.a.  \\ \cline{2-5}
& Shibata et al.~\cite{Shibata2025}   & 4   & 0.812  & n.a.  \\ \cline{2-5}
& \multirow{2}{*}{Our Approach}       & 4   & $0.930 \pm 0.086$ & $0.934 \pm 0.084$  \\  \cline{3-5}
&                                     & 6   & $0.938 \pm 0.075$ & $0.940 \pm 0.075$  \\\hline\hline

\multirow{4}{*}{AOS} & 
Yoshida et al.~\cite{Yoshida20232}    & 4   & 0.90  & n.a.  \\ \cline{2-5}
& Shibata et al.~\cite{Shibata2025}   & 4   & 0.930  & n.a.  \\ \cline{2-5}
& \multirow{2}{*}{Our Approach}       & 4   & $0.940 \pm 0.081$ & $0.941 \pm 0.081$  \\  \cline{3-5}
&                                     & 6   & $0.941 \pm 0.063$ & $0.944 \pm 0.066$  \\\hline

\end{tabular}
}
\end{table}

Table~\ref{tab:comparison-ford} summarizes the performance of our best-accuracy configuration on our self-recorded dataset for the three different data splitting methods and both gesture sets (four and six gestures). 
Applying~\textit{PS}, the model achieves very high accuracy and precision (up to~\(0.941\) and~\(0.961\)), indicating reliable within-subject classification.
In the more challenging~\textit{LOSO} setting, accuracy remains above~\(0.93\) for both gesture sets, showing good generalization to unseen participants.
Adding a single calibration session from the target user in the~\textit{AOS} setting further improves accuracy to~\(0.94\) (four~gestures) and~\(0.941\) (six~gestures), suggesting that a small amount of user-specific data is sufficient to stabilize performance across users. Here, the gestures swipe-down and tap achieve highest accuracy of~\(0.96\) and swipe-right has lowest accuracy of~\(0.91\).

Notably, performance for six gestures is comparable to that for four gestures across all three data splitting methods, indicating that extending the gesture set does not substantially degrade recognition performance on our self-recorded dataset.
\subsection{Comparison with Related Work}
Table~\ref{tab:comparison-ford} compares the best-performing configuration on our self-recorded dataset with the original~\SmaTable system~\cite{Yoshida20231, Yoshida20232} and Shibata et al.~\cite{Shibata2025} on the original \SmaTable dataset. For comparability, we focus on the four swipe gestures under \textit{PS}, \textit{LOSO}, and \textit{AOS}. 
The model by Shibata et al.~\cite{Shibata2025} is heavily optimized for FPGA deployment, using a compressed and integer-quantized architecture tailored to a specific accelerator design, whereas our model is trained in full precision. The comparison should therefore be interpreted at the system level rather than as a direct architecture-level comparison.

Using the \textit{PS} strategy, our approach achieves an accuracy of~\(0.941\), slightly above the original \SmaTable result~(\(0.92\)) and close to the FPGA-optimized model~(\(0.97\))~\cite{Shibata2025}. In the \textit{LOSO} setting,  our approach reaches~\(0.930\), clearly outperforming the original system~(\(0.670\)) and the FPGA implementation (\(0.812\)), corresponding to drops of~\SI{25}{\%} and~\SI{16}{\%} compared to their PS performance. In the \textit{AOS} setting, the \SmaTable baseline attains \(0.93\) accuracy, while our approach achieves~\(0.94\) for~four~gestures and~\(0.941\) for~six~gestures. 

A key difference to the \SmaTable results is the much smaller \textit{LOSO} and \textit{AOS} gap on our dataset, which we attribute to the larger number of participants and thus more diverse \textit{LOSO} training sets, so that a single calibration session adds only limited benefit.
\section{Conclusion and Future Work}
\label{sec:conclusion_future_work}
This paper presents an approach for optimizing an end-to-end signal processing and classification pipeline for vibration-based gesture recognition on furniture. We designed a low-noise custom sensor hardware and a configurable data-to-model pipeline combining pre-processing blocks with depthwise separable 1D-CNNs. A systematic configuration search on a self-recorded dataset with~15~participants and six gestures yielded the best configuration with an accuracy of~\SI{97.32}{\%} and in total~8,722 model parameters. Compared to related work, our approach includes a strong user-independent performance under leave-one-subject-out evaluation.

In future, we plan to enhance the dataset with new table types, table materials, more sensor channels, new gesture vocabularies. Also, we aim to transfer the end-to-end signal processing pipeline into the hardware for real-time prediction by building a hardware-related code-generator of the pre-processing methods and enhance the workflow by considering model compression for efficient embedded implementation. 

\bibliographystyle{IEEEtran}
\bibliography{reference}
\end{document}